\documentclass[preprint,12pt]{elsarticle}

\def\lbldef#1#2{\expandafter\gdef\csname #1\endcsname {#2}}

\def\href#1#2{#2}

\usepackage{color}
\usepackage{graphicx}
\usepackage{amssymb}
\journal{Physics of the Dark Universe}

\begin{document}
	
	\begin{frontmatter}
		
		\title{A Resolution of the Monopole Problem in the $R_{\rm h}=ct$ Universe} 
		
		\author[1]{Fulvio Melia\footnote{John Woodruff Simpson
				Fellow. E-mail: fmelia@email.arizona.edu}} 
		
		\address[1]{Department of Physics, The Applied Math Program, and Department of Astronomy,
			The University of Arizona, AZ 85721, USA}
		
		\begin{abstract}
		Spontaneous symmetry breaking in grand unified theories is thought to have produced
an exceedingly large number of magnetic monopoles in the early Universe. In the absence
of suppression or annihilation, these very massive particles should be dominating the
cosmic energy budget today, but none has ever been found. Inflation was invented in
part to dilute their number, thereby rendering their density undetectable by current
instruments. Should the inflationary paradigm not survive, however, the ensuing
disagreement between theory and observation would constitute a cosmological `monopole
problem' and create further tension for any extension to the standard model of particle
physics. But as is also true for all horizon problems, a monopole overabundance emerges
only in cosmologies with an initial period of deceleration. We show that the alternative
Friedmann-Lema\^itre-Robertson-Walker cosmology known as the $R_{\rm h}=ct$ universe
completely eliminates all such anomalies rather trivially and naturally, without the
need for an inflated expansion. We find that the monopole energy density today would be
completely undetectable in $R_{\rm h}=ct$. Evidence continues to grow that the zero
active mass condition from general relativity ought to be an essential ingredient in
$\Lambda$CDM.
\end{abstract}
		
	\end{frontmatter}
	
\section{Introduction}\label{intro}
At the core of the standard model of particle physics is the unification
of electromagnetism and the weak nuclear force into a single SU(2) $\times$ U(1)
symmetry \cite{Glashow:1961,Salam:1964,Weinberg:1967}, which is broken at low
energies because of a Higgs field \cite{Englert:1964,Higgs:1964,Guralnik:1964}
that acquired a vacuum expectation value \cite{Atlas:2012} some $10^{-11}$
seconds after the Big Bang \cite{Melia:2018a}.

But there are many theoretical reasons to expect an even bigger---or grand---unification
of all known elementary particle forces (except gravity), into a so-called Grand Unified
Theory (GUT), as first proposed by Georgi and Glashow in 1974 \cite{Georgi:1974}. These
include: (i) the fact that the complex structure of particles in the standard model,
comprising three different gauge symmetries and a wide assortment of particle
properties, is more simply explained within a grand unified scheme; and (ii) that the
interaction strengths are not fixed constants. They vary with energy in such a way
that they meet when extrapolated to $\sim 10^{16}$ GeV \cite{Patrignani:2016}.

Georgi and Glashow achieved this unification quite simply using only the SU(5)
non-Abelian gauge symmetry and one additional Higgs scalar field. If this second
Higgs field also gained a vacuum expectation value---presumably at a temperature
$kT\sim 10^{15-16}$ GeV---the GUT symmetry would have been broken into the SU(2) and
U(1) symmetries of the electroweak theory and a separate SU(3) symmetry describing
the strong nuclear force.

Since then, other GUTs have been proposed and today the SU(5) version is not
unique in describing all strong and electroweak forces using one grand unified
scheme, relying instead on different gauge groups, such as SO(10) \cite{Fritzsch:1975}
and E(6) \cite{Gursey:1976}. Nevertheless, all of them have one key property in
common---the creation of 't Hooft-Polyakov monopoles when the GUT symmetry is
spontaneously broken \cite{tHooft:1974,Polyakov:1974,Preskill:1979}. This happens
because whenever a U(1) symmetry remains after a gauge symmetry is broken by a
Higgs field, so-called stable `hedgehog' Higgs configurations also persist
\cite{Kibble:1976,Preskill:1984}. But a U(1) gauge symmetry is required for
electromagnetism, so any GUT that unifies all three forces other than gravity
must always have stable `hedgehogs,' i.e., 't Hooft-Polyakov monopoles.

These features are examples of a topological defect, or a topological soliton
\cite{Manton:2004}, and are {\it extremely stable}, given that they cannot be
turned continuously into the uniform vacuum state. GUT monopoles might
therefore have created a problem for cosmology when the GUT symmetry was
broken if the Universe began with a temperature $kT\gg 10^{16}$ GeV and cooled
soon after the Big Bang. They would have appeared physically as quanta of energy
localized within small volumes, interacting as massive particles with their
environment.

As we shall see in \S~\ref{defects}, a monopole problem arises in standard
Big Bang cosmology because even simple calculations suggest that so many of
them would have been created that the cosmic dynamics today would be
completely overwhelmed by their gravitational influence, which is clearly
not the case. The earliest estimates assumed that the monopole density was
at some time in thermal equilibrium \cite{Zeldovich:1978,Preskill:1979}, but
even other approaches avoiding this assumption \cite{Guth:1980,Einhorn:1980}
arrived at similar conclusions.

The favored explanation for the absence of a GUT monopole presence today is
inflation, introduced in various guises and for a variety of reasons
\cite{Starobinsky:1979,Kazanas:1980,Linde:1982}, but specifically to solve
the monopole, horizon and flatness problems by Alan Guth in 1981 \cite{Guth:1981}.
In this picture, the early Universe was dominated by a scalar field with a potential
producing a de Sitter type of accelerated expansion which simply diluted the
monopole density away to insignificant levels, before resettling back to its
standard expansion driven by matter, radiation and (as we now know) an unknown
form of dark energy. In order for this work, the inflationary expansion had
to occur after the creation of the GUT monopoles (or during it, as originally
thought).

But even after three decades of development, we still do not have a complete
picture of how inflation is supposed to have worked. Many would categorize it
as more of a general idea than a specific, well-understood theory. In recent
years, there have been several indications that its foundational tenets are
simply not consistent with the data. For example, in light of the latest
{\it Planck} measurements \cite{PlanckVI:2020}, significant doubts have
been raised concerning its assumed initial conditions \cite{Ijjas:2014}.

Even more compellingly, a careful re-analysis of the temperature anisotropies
in the CMB has shown to a high degree of confidence that the primordial power
spectrum $P(k)$ has a hard cutoff, $k_{\rm min}=(3.14\pm0.36)\times 10^{-4}\;{\rm Mpc}^{-1}$
\cite{MeliaLopez:2018,Melia:2021b,Sanchis-Lozano:2022}, which creates significant
tension with all slow-roll inflationary potentials. In order to correctly
account for $P(k)$, the inflationary potential must dominate over the
field's kinetic energy, but then the presence of $k_{\rm min}$ inhibits
inflation from producing an expansion with a sufficient number of e-folds
to simultaneously solve the horizon problem \cite{LiuMelia:2020}.

Enough concerns have now been leveled at the general inflationary paradigm that
we should seriously consider whether it could have happened at all. At this point,
we could either resort to a drastic simplification and merely claim that
monopoles were never produced, which would then shift the problem to the
standard model of particle physics, or instead seek an alternative cosmology
in which the monopole problem never emerges.

Indeed, we shall demonstrate in this paper that the alternative
Friedmann-Lema\^itre-Robertson-Walker (FLRW) cosmology, known as the
$R_{\rm h}=ct$ universe \cite{MeliaShevchuk:2012,Melia:2020}, trivially
and {\it naturally} eliminates the GUT monopole problem completely. And
it does so while also obviating, or significantly mitigating, all the other
conflicts and inconsistencies in $\Lambda$CDM (see, e.g., Table~2 in
ref.~\cite{Melia:2018e} and the more complete discussion in ref.~\cite{Melia:2020}),
including all horizon problems that plague the standard model, to which
the monopole anomaly is closely related.

\section{Topological Defects and the Monopole Problem}\label{defects}
We have no reason to believe that the temperature in the Universe was
initially lower than $10^{16}$ GeV at the Big Bang. It is therefore reasonable
to assume that the Universe must have undergone a phase transition as
it cooled, corresponding to a critical temperature $kT_{\rm c}$ of order
the unification energy scale. Above $T_{\rm c}$, the Higgs scalar field,
which acts as an order parameter for the symmetry breaking, had zero
expectation value, so no monopoles were present. They formed once $T$ dropped
below $T_{\rm c}$, however, and were extremely stable
\cite{Kirzhnits:1972,Weinberg:1974,Dolan:1974}.

The basic idea behind the formation of GUT monopoles follows the approach
first taken by Kibble in 1976 \cite{Kibble:1976}, and later generalized
to include its applicability to grand unified theories \cite{Preskill:1984,Rajantie:2003a}.
The detailed mechanism depends on whether the phase transition was
second-order (or weakly first-order), producing large fluctuations,
or whether it was strongly first-order, with an associated supercooling
progression. The expected monopole density, however, is very similar in
both cases.

At a fundamental level, one can reasonably argue that the Higgs field during
the phase transition could not have been correlated over distances greater
than some characteristic scale $\xi$. This is easy to understand in the
context of general relativity, given that the transition would have occurred
over a finite time, so the Higgs field at two spacetime points separated beyond
their causal limit would have settled into vacuum expectation values independently
of each other, producing a domain structure with a characteristic size $\xi$
for each domain.

But the Higgs field had to be continuous, so it would have interpolated
smoothly between two adjacent domains. Nevertheless, there is a probability
$p$ (not much smaller than 1) that the scalar field orientation ended up being
topologically nontrivial at the intersection point of several independent
domains, so a monopole or antimonopole would have formed there. One therefore
estimates that the density of GUT monopoles created at the phase transition
must have been
\begin{equation}
n_{\rm m}(t_{\rm GUT})\sim p\xi^{-3}\;,\label{eq:nGUT}
\end{equation}
with $p\sim 1/10$ in typical grand unified theories \cite{Preskill:1984}.

At least in the case of a second-order (or weakly first-order) phase
transition, $\xi$ might have become very large as $T$ approached $T_{\rm c}$,
but even for this situation the domain structure of the Higgs field could not
have violated the causality limit \cite{Guth:1980,Einhorn:1980}. The largest
correlation length one could contemplate is thus the gravitational (or
Hubble) radius,
\begin{equation}
R_{\rm h}(t_{\rm GUT})\equiv \frac{c}{H(t_{\rm GUT})}\;,\label{eq:RhGUT}
\end{equation}
where $H(t_{\rm GUT})$ is the Hubble parameter at the time the phase transition
took place (see ref.~\cite{Melia:2018b} for a discussion of causal limits based on
the gravitational horizon in cosmology). And so we conclude from this that,
in a second-order (or weakly first-order) transition, the initial GUT monopole
density in the comoving frame would have been set by the constraint
$\xi\le R_{\rm h}(t_{\rm GUT})$, implying that
\begin{equation}
n_{\rm m}(t_{\rm GUT})\ge pR_{\rm h}(t_{\rm GUT})^{-3}\;.\label{eq:nGUT2}
\end{equation}

If the GUT transition was first-order, the emergence of a vacuum
expectation value for the Higgs field would have proceeded via the nucleation
of stable bubbles within an unstable, unbroken medium. These bubbles would
have expanded at lightspeed to fill the Universe with the stable phase
\cite{Coleman:1977}. GUT monopoles would then have been created where the
bubbles collided and merged, based on the same Kibble mechanism described
above. The principal difference between these two outcomes would have been
the temperature at which the transition was finalized, which was presumably
$T_{\rm c}$ in the former, but somewhat lower in the latter, given that
the phase with unbroken symmetry would have persisted until the bubble
nucleation was complete. A somewhat lower temperature for the first-order
transition implies a later time, so the predicted density of monopoles
in this case would have been
\begin{equation}
n_{\rm m}(t_{\rm bubble})\ge pR_{\rm h}(t_{\rm bubble})^{-3}\;,\label{eq:nGUT3}
\end{equation}
with $t_{\rm bubble}> t_{\rm GUT}$. Such differences, however, have no
material impact on our discussion, so we shall henceforth simply assume that
the initial GUT monopole density was generically given by Equation~(\ref{eq:nGUT2}).

Once created, the abundance of monopole-antimonopole pairs might have been
reduced by annihilations, but this process would have been very inefficient
due to the very large monopole mass and the rapid cosmic expansion. As such,
the density of GUT monopoles in the comoving frame would have been hardly
reduced at all below the estimate given in Equation~(\ref{eq:nGUT2})
\cite{Zeldovich:1978,Preskill:1979,Dicus:1982,Preskill:1984}.

In anticipation of our proposed solution to the monopole anomaly, let us
now examine why this monopole abundance creates a significant problem for
standard Big Bang cosmology. Adopting the flat $\Lambda$CDM model, we may set
the expansion factor $a(t)$ equal to $1$ today. Then, with $kT_{\rm c}=
10^{16}$ GeV, the equation
\begin{equation}
T_{\rm c}=(1+z_{\rm GUT})T_0\;,\label{eq:Tc}
\end{equation}
in terms of today's CMB temperature, $T_0=2.72548\pm0.00057$ K, yields
the redshift, $z_{\rm GUT}\sim 4.3\times 10^{28}$, at which the GUT phase
transition would have taken place. The corresponding expansion factor was
\begin{equation}
a(t_{\rm GUT})=\frac{1}{1+z_{\rm GUT}}\sim 2.3\times 10^{-29}\;.\label{eq:atGUT}
\end{equation}

In the standard model, the Hubble parameter may be written
\begin{equation}
H(a) = H_0\sqrt{\Omega_{\rm m}a^{-3}+\Omega_{\rm r}a^{-4}+\Omega_\Lambda}\;,\label{eq:Ha}
\end{equation}
which we evaluate using the {\it Planck} optimized parameters, including the
Hubble constant $H_0=67.4\pm0.5$ km s$^{-1}$ Mpc$^{-1}$, and normalized densities
$\Omega_{\rm m}=0.315\pm 0.007$ (matter), $\Omega_{\rm b}=0.0377\pm 0.0002$ (baryons),
$\Omega_{\rm r}=5.370\pm0.001\times 10^{-5}$ (radiation) and $\Omega_\Lambda=0.685\pm0.015$
(cosmological constant) \cite{PlanckVI:2020}. Thus, we find that the gravitational
(or Hubble) radius at the GUT scale (Eq.~\ref{eq:RhGUT}) must have been
\begin{equation}
R_{\rm h}(t_{\rm GUT})\sim 3.9\times 10^{-28}\;{\rm cm}\;.\label{eq:RhGUT2}
\end{equation}

If we now assume that the comoving density of GUT monopoles has remained more
or less constant for $t\ge t_{\rm GUT}$, we find that their physical density
today would be
\begin{equation}
n_{\rm m}(t_0)\sim p\left[\frac{a(t_{\rm GUT})}{R_{\rm h}(t_{\rm GUT})}\right]^3\;,\label{eq:nt0}
\end{equation}
which yields
\begin{equation}
n_{\rm m}(t_0)\sim 21\left(\frac{p}{0.1}\right)\;{\rm m}^{-3}\;.\label{eq:nt02}
\end{equation}
By comparison, the proton density is estimated to be
\begin{equation}
n_{\rm H}(t_0)\approx \frac{\Omega_{\rm b}\rho_{\rm c}}{m_{\rm H}}\sim
0.2\;{\rm m}^{-3}\;,\label{eq:nH}
\end{equation}
where $\rho_{\rm c}\equiv 3H_0^2/8\pi G$ is the critical mass density and
$m_{\rm H}$ is the mass of the hydrogen atom.

A comparison of Equations~(\ref{eq:nt02}) and (\ref{eq:nH}) shows clearly
why we have a monopole problem in cosmology, since the magnetic monopole
density would not only be much larger than that of baryons today but, when
coupled with their enormous mass difference (i.e., a factor of $\sim 10^{16}$),
implies that the monopole contribution to the energy density of the Universe
would be ridiculously high, certainly well beyond any reasonable upper limit
placed by ongoing searches for these topological defects
\cite{Preskill:1984,Rajantie:2012}.

A careful inspection of the argument leading up to Equation~(\ref{eq:nt02})
would reveal that the standard model fails because it predicts a decelerated
expansion at early times (i.e., $\dot{a}\sim t^{-1/2}$; see Eq.~\ref{eq:Ha}
with $H(a)\equiv\dot{a}/a$), greatly inhibiting the rate at which the volume
per monopole grew in comparison with the size of the visible Universe, i.e.,
$\dot{R}_{\rm h}$. The monopole problem is thus caused by the same flaw that
gives rise to the various horizon problems in standard cosmology
\cite{Melia:2013c,Melia:2018a}.

The accelerated spurt during inflation is supposed to have overcome this
deficiency by greatly expanding the physical volume per monopole and thereby
hugely diluting their density to undetectable levels today. But if inflation
were to eventually go away, as some of the observations are now suggesting, we
would be left with a significant conflict between the standard model of particle
physics and the current standard model of cosmology. In the next section, we
shall demonstrate how this impasse---like all the other current problems in
$\Lambda$CDM---is completely and {\it naturally} removed by the alternative FLRW
cosmology known as the $R_{\rm h}=ct$ universe.

\section{The $R_{\rm h}=ct$ universe}\label{Rh}
The $R_{\rm h}=ct$ cosmology has been under development for over fifteen years
\cite{Melia:2003,MeliaShevchuk:2012,Melia:2020}. As of today, more than 27
different kinds of observation have been used in comparative studies between
this model and $\Lambda$CDM, at both high and low redshifts, employing a broad
range of integrated and differential measures, such as the luminosity and angular
diameter distances, the redshift-dependent expansion rate, and the redshift-age
relationship.  In all of the tests completed and published thus far, $R_{\rm h}=ct$
has accounted for the data at least as well as the standard model, and often
much better. A recent compilation of the papers reporting this work may
be found in Table~2 of ref.~\cite{Melia:2018e}. A more complete description
of this model, including its foundational underpinnings, may be found in
\cite{Melia:2020}.

Briefly, the original motivation for this model was the
emergence of a very unusual `coincidence' in the cosmological data,
suggesting that the apparent (or gravitational) horizon in the
Universe was equal to the light-travel distance since the Big Bang
\cite{Melia:2003,Melia:2007,Melia:2018b}. It is very straightforward to convince
oneself that, given the various periods of acceleration and deceleration
in the standard model, this equality can only happen once in the entire
history of the Universe, and yet it is happening right now, at time
$t_0$, just when we happen to be looking \cite{MeliaShevchuk:2012}. Of 
course, the probability for such a chance coincidence is thus 
effectively zero if $\Lambda$CDM is the correct cosmology. It is well
known that the standard model suffers from several inexplicable
coincidences, but this one is arguably the worst.

The simplest `solution' to this conundrum is that the equality
$R_{\rm h}=ct$ (hence the eponymous name for the model) must
be true at all times, not just at this instant. Then it wouldn't
matter when the observations are made, since the same condition
would be valid at all times $t$ smaller than, or larger than, $t_0$.

This equality, however, implies (via the Friedmann equations) that
the cosmos expands at a constant rate, with an expansion factor
$a(t)=t/t_0$. This contrasts with the variable expansion rate
predicted by $\Lambda$CDM, so the earliest work with this hypothesis 
has revolved around the acquisition of empirical evidence supporting 
this unexpected scenario. Needless to say, the general degree of
success enjoyed by the standard model over the past few decades makes
it difficult to believe that the history of the Universe could be
adequately accounted for by such a different paradigm. And yet,
test after test, now including over 27 different types of data, at
both high and low redshifts, based on measurements of the luminosity
distance, or the angular diameter distance, or the redshift-dependent
Hubble expansion rate or (perhaps most spectacularly) the redshift-time
dependence, have all shown that the observations quite compellingly
favor $R_{\rm h}=ct$ over the standard model. A quick perusal of
the aforementioned Table~2 in ref.~\cite{Melia:2018e} would show
that the `score' is effectively 27 to 0 in favor of the former model.

And this compilation does not include the most recent comparative
tests based on the latest JWST observations 
\cite{Harikane:2022,Donnan:2022,Naidu:2022a,Bradley:2022} showing that
the timeline for the formation of structure predicted by $\Lambda$CDM
in the early Universe is strongly disfavored, while that predicted by
$R_{\rm h}=ct$ matches the data almost exactly \cite{Melia:2023b,Melia:2023c}.

The successful empirical support it has received from this body of
observational work has motivated a deeper exploration of its origin and 
viability. As we now understand it, $R_{\rm h}=ct$ is essentially $\Lambda$CDM, 
but with a critical added constraint to its total equation-of-state, known
as the zero active mass condition in general relativity, i.e., $\rho+3p=0$, 
where $\rho$ and $p$ are, respectively, the total energy density and pressure 
in the cosmic fluid. Again, it is straightfoward to see this from
the Friedmann equations when one imposes the constraint $R_{\rm h}=ct$,
which in turn implies that $a(t)=t/t_0$.

But more recent theoretical work appears to show that this condition 
may be necessary for the proper usage of the FLRW metric in a cosmic 
setting \cite{Melia:2022b,Melia:2023}. Evidence is growing that the
choice of lapse function, $g_{tt}=1$, in the FLRW metric precludes
any possibility of an accelerated expansion, given that it permits
no time dilation in the accelerated frame relative to a local
free-falling frame. This is still work in progress, awaiting further
independent confirmation. If correct, this fundamental underpinning
explains why the zero active mass equation-of-state must produce an 
expansion profile with $a(t)=(t/t_0)$ at all redshifts, including 
the early Universe, where the monopole problem emerges.

The theoretical support this model now receives is extensive, impacting every
area in which $\Lambda$CDM has a major problem or inconsistency. For example,
the $R_{\rm h}=ct$ universe completely eliminates the CMB temperature
\cite{Melia:2013c} and electroweak \cite{Melia:2018a} horizon problems, 
without the need for inflation. It solves
the cosmic entropy problem \cite{Melia:2021d}, and provides a natural explanation
for the origin of the cutoff $k_{\rm min}$ in the primordial power spectrum
\cite{Melia:2019b}. It also completely and naturally removes the time compression
problem in $\Lambda$CDM, in which galaxies \cite{Melia:2014a} and quasars
\cite{Melia:2013b} would otherwise appear far too early in its history. Quite
remarkably, the $R_{\rm h}=ct$ cosmology even explains the origin of rest mass
energy \cite{Melia:2021a}. In addition to these notable successes, several other
applications and conflict resolutions have also been reported in both the
primary and secondary literature.

In this paper, we address one of the few remaining topics yet to be
broached in this comparative study---i.e., the third and final original motivation
for the introduction of inflation back in the early 1980's. The monopole problem
has been invoked on countless occasions as important phenomenological support
for the existence of an inflaton scalar field in the early Universe. But as we
have noted elsewhere in this paper, the observations now appear to be
retreating from this paradigm, creating a growing schism between the current
standard model of particle physics and $\Lambda$CDM. In the next section, however,
we shall demonstrate how the $R_{\rm h}=ct$ universe completely removes the
monopole problem naturally and elegantly, adding to its long list of
accomplishments discussed above.

\section{GUT Monopoles in the $R_{\rm h}=ct$ universe}\label{GUTmonopoles}
As noted earlier in \S~\ref{defects}, horizon problems emerge only in
cosmologies with an early period of decelerated expansion, such as
$\Lambda$CDM \cite{Melia:2013c}. Horizon problems, which are closely
related to an overabundance of magnetic monopoles, therefore do not
even emerge in a model such as $R_{\rm h}=ct$, whose expansion
never decelerated.

This is very easy to demonstrate quantitatively. In this alternative
cosmology, we have $a(t)=t/t_0$ and $R_{\rm h}(t)=ct$, so the monopole
density in Equation~(\ref{eq:nt0}) simply becomes
\begin{equation}
n_{\rm m}(t_0)\sim \frac{p}{R_{\rm h}(t_0)^3}\;,\label{eq:nRh}
\end{equation}
regardless of when the GUT phase transition occurred. In other words,
since the initial density of magnetic monopoles in grand unified theories
was expected to be of order $p$ per Hubble volume, it would have remained
of order $p$ per Hubble volume throughout history, including today. Needless
to say, this density is completely undetectable, given that
$R_{\rm h}(t_0)\sim 1.4\times 10^{28}$ cm, and monopoles would have zero
influence on the expansion dynamics.

Thus, the $R_{\rm h}=ct$ universe trivially solves the horizon and monopole
problems because the observable Universe today was always causally connected
from the very beginning. Whatever density of monopoles was created at the
GUT transition would have remained constant in time because the comoving
volume filling the Hubble sphere at $t_{\rm GUT}$ in this cosmology would
also fill the entire visible Universe today.

\section{Conclusion}\label{conclusion}
With a resolution of the monopole problem we have discussed in this
paper, all of the difficulties inflation was designed to overcome
have now been eliminated in the context of $R_{\rm h}=ct$. This model
not only accounts for the data at both low and high redshifts generally
better than $\Lambda$CDM, it also obviates the need for additional exotic
mechanisms to solve problems that may not have been real to begin with.

Looking to the future, several observational campaigns will directly test
the predictions of $\Lambda$CDM versus $R_{\rm h}=ct$, promising to
unambiguously reject one or the other (or perhaps both) of these models.
Chief among them will be the measurement of redshift drift \cite{Sandage:1962}---a
clear determination of whether the cosmic expansion is accelerating or not.
This effect merely requires the validity of the cosmological principle,
and produces a temporal change in the redshift of fixed comoving sources
if the expansion of the Universe is variable
\cite{Corasaniti:2007,Quercellini:2012,Martins:2016}.
The Extremely Large Telescope high resolution spectrometer (ELT-HIRES)
\cite{Liske:2014} will facilitate measurements in the redshift range
$2\le z \le 5$, while the Square Kilometer Phase 2 Array (SKA)
\cite{Kloeckner:2015} will do the same for $z\le 1$. Complementary
observations may also be feasible with 21 cm experiments, e.g.,
the Canadian Hydrogen Intensity Mapping Experiment (CHIME)
\cite{Yu:2014}. The $R_{\rm h}=ct$ universe predicts zero drift at all
redshifts \cite{Melia:2016b,Melia:2022d}, so the required outcome of
these campaigns is essentially just a yes/no answer, which might be
achievable with a baseline of just five to ten years.

This is but one of many new and improved measurements that should
revolutionize our view of cosmic history. Hopefully, the work we have
reported in this paper will help to set the stage for the
meaningful interpretation of the new data, and help us to clearly
identify the correct cosmological model.

\section*{Acknowledgements}
I am grateful to Amherst College for its support through a John Woodruff
Simpson Lectureship. I am also grateful to the anonymous
referee for their careful, expert reading of this manuscript.

\bibliographystyle{elsarticle-harv}
\bibliography{ms}

\end{document}